\documentclass[twoside]{article}

\usepackage{PRIMEarxiv}

\usepackage[utf8]{inputenc} 
\usepackage[T1]{fontenc}    
\usepackage{hyperref}       
\usepackage{url}            
\usepackage{booktabs}       
\usepackage{amsfonts}       
\usepackage{nicefrac}       
\usepackage{microtype}      
\usepackage{lipsum}
\usepackage{fancyhdr}       
\usepackage{graphicx}       
\graphicspath{{media/}}     
\usepackage{subcaption}

\title{An open-source Autonomous Surface Vehicle for Acoustic Tracking, Bathymetric and Photogrammetric Surveys}

\author{
  Pierre Gogendeau \\
  IFREMER DOI\\
  Le Port, 97420 Reunion \\
  \texttt{pierre.gogendeau@gmail.com} 
  \AND
  Sylvain Bonhommeau \\
  IFREMER DOI\\
  Le Port, 97420 Reunion \\
  \texttt{sylvain.bonhommeau@ifremer.fr} \\
  \And
  Hassen Fourati \\
  GIPSA-Lab/U. Grenoble Alpes/CNRS/Inria/INP* \\
  38000 Grenoble, France \\
  \texttt{hassen.fourati@grenoble-inp.fr} \\
  \And
  Mohan Julien \\
  IFREMER DOI\\
  Le Port, 97420 Reunion \\
  \texttt{mohan.julien@ifremer.fr} \\
  \And
  Matteo Contini\\
  IFREMER DOI\\
  Le Port, 97420 Reunion \\
  \texttt{Matteo.Contini@ifremer.fr} \\
  \And
  Thomas Chevrier \\
  COOOL SAS \\
  Saint-Leu, 97436 Reunion \\
  \texttt{t.chevrier.cooolresearch@gmail.com} \\
  \And
  Anne Elise Nieblas\\
  COOOL SAS \\
  Saint-Leu, 97436 Reunion \\
  \texttt{coool.research@gmail.com} \\
  \And
  Serge Bernard\\
  LIRMM/CNRS University of Montpellier\\
  Montpellier, 34000 France \\
  \texttt{serge.bernard@cnrs.fr} \\
}

\begin{document}
\maketitle
\thanks{\textit{\underline{Citation}}: \textbf{submitted to  Ocean Engineering (Elseiver), June 2024}} 

\begin{abstract}

Autonomous Surface Vehicles (ASV) are becoming more affordable and include a wide variety of sensors and capacities with applications from ocean physics such as the Saildrone project to ecology with the tracking of marine species in the wild. Here, we present a multi-modal, affordable, open source, and reproducible ASV to track marine animal in shallow waters, collect information on bathymetry, and carry out photogrammetry surveys. The current specification enables scientists to track an animal equipped with an acoustic tag for 5~h and a spatial accuracy of 1~m. For bathymetric or photogrammetry surveys, the ASV can cover 100 x 100~m areas in 2~h with a distance of 1-m between transects. Depending on the sensors included in the ASV, it has a price ranging from \$2,434 to \$11,072. We illustrate these developments with a case study and a field survey for each of the different application proposed.
\end{abstract}




\section{Introduction} 
\label{sec:introduction}

Most of USV/ASV are very expensive and reserved for the military \cite{liu_unmanned_2016,yan_development_2010}, the industry \cite{noauthor_surfbee_2021}, or some scientific institutes \cite{kimball_whoi_2014,ferri_hydronet_2015}. In the past few years, several projects emerged, proposing small and low-cost ASV/USV under \$5000 without specific sensors \cite{manda_low_2015,raber_low-cost_2019,lambert_low-cost_2020}. This accessibility improvement is made thanks to some companies proposing  cheap, reliable and open-source electronics and marine robotic parts. For instance, the T200 thruster made by \textit{Blue Robotics} is used by many hobbyists \cite{noauthor_boogie_nodate}, scientists \cite{martorell_xiroi_2018}, and industrial \cite{noauthor_surfbee_2021} projects. 
We found the same evolution in software programs. Professionals and hobbyists developed good quality, easy to use, well documented, and open-source autopilots systems. For example, $Ardupilot$ is now embedded in various vehicles such as drones, rovers, remotely operated vehicle (ROV) and boats \cite{zhao_design_2020,he_ard-mu-copter_2015,burke_4g-connected_2020}. It can also be used for data logging, analysis, and simulation. The open community linked to these projects makes them in constant and dynamic evolution.  \\

\begin{figure*}[ht]
\centering
\includegraphics[width=\textwidth]{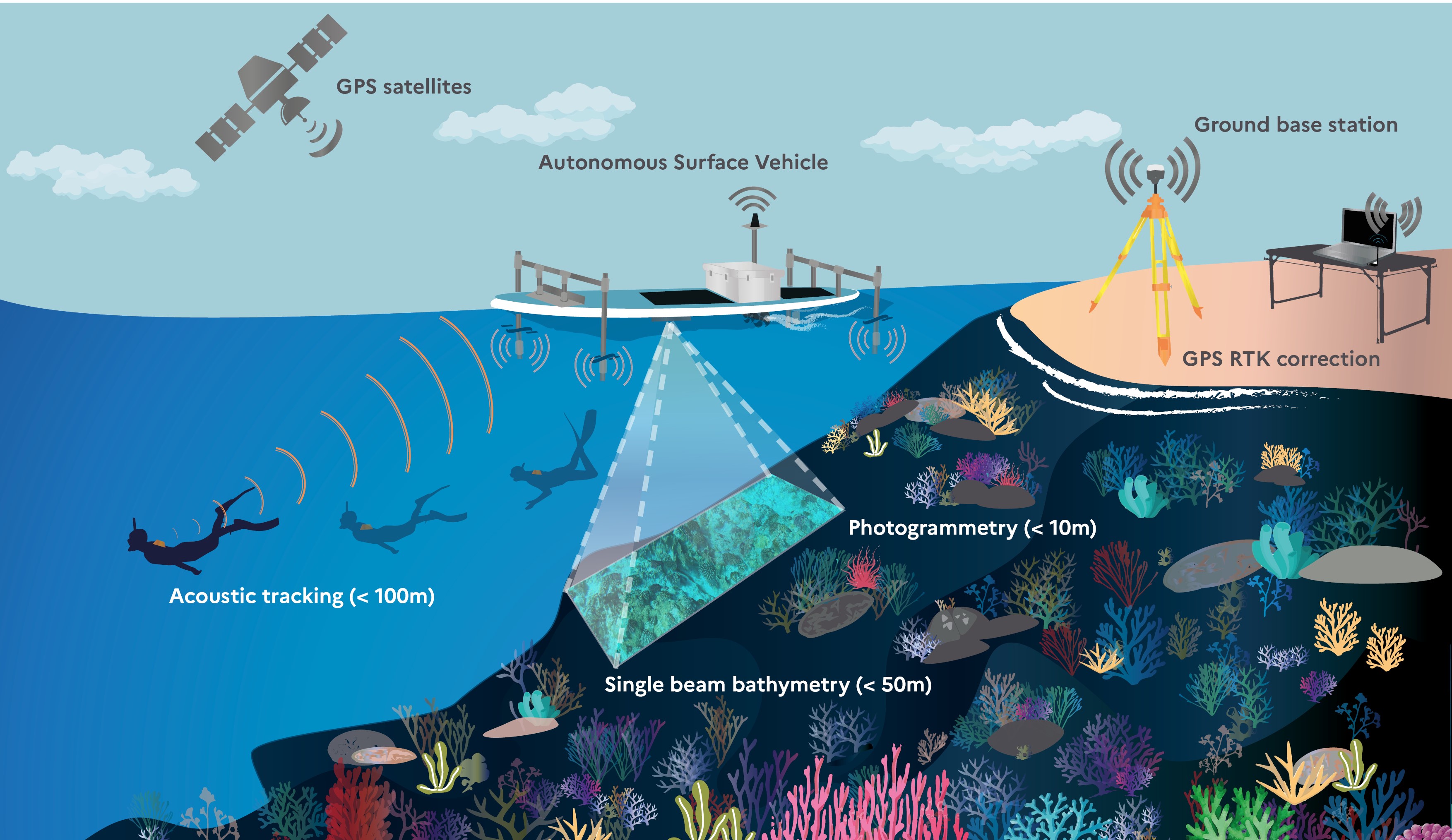}
\caption{Schematic of the functioning of the $Plancha$ autonomous surface vehicle (ASV) and the data collected during an ASV survey: Autonomous acoustic tracking, single-beam bathymetric survey, and photogrammetric survey}
\label{fig4}
\end{figure*}
%

\noindent In this paper, we present an ASV developed for 3 main missions: 

\begin{itemize}
    \item{Autonomous acoustic tracking of a marine animal}
    \item{Bathymetric survey}
    \item{Photogrammetric survey}
\end{itemize}

\subsection{Autonomous acoustic tracking of a marine animal}

Habitats use by marine species and their behavior can be adressed with advances in biologging technology that enable fine-scale geolocated trajectories. Biologging refers to the deployment of autonomous devices onto free-ranging animals to collect physical and biological information through its different sensors \cite{ropert-coudert_bio-logging_2012}. A common technique for geolocated trajectory estimation of marine animals is dead-reckoning, using a fusion of inertial data, sensor speed, and GPS positions \cite{wilson_all_2007}. However, DR induce high computing effort \cite{gunner_dead-reckoning_2021}.

Underwater fine-scale geolocated tracking is also possible with acoustic systems composed of transmitters and receivers. Some systems with anchored or buoy receivers need dense acoustic antenna arrays \cite{espinoza_testing_2011} to use their geolocation algorithms. These systems are not adapted for some marine species because the distance they cover per day can be several kilometers. Some other acoustic systems are more compact, like ultra-short baseline (USBL) and short baseline (SBL) acoustic systems. USBL and SBL calculate the range of an acoustic transponder based on the signal time of arrival (TOA) or time of flight (TOF). In addition, USBL uses a phase-differencing algorithm with the receiver baseline to get the bearing angle \cite{paull_auv_2014}. With the calculated relative position, both systems infer the geolocated position of the transponder adding the global navigation satellite system (GNSS) position of the receiver system. USBL receiver systems are more compact and offer a better range and accuracy. For the Blue Print USBL \footnotemark \footnotetext{\url{https://www.blueprintsubsea.com/seatrac/seatrac-lightweight}}, the range is 1 km with 0.1 m accuracy compared to the 100 m with 1 m accuracy of the Waterlinked UGPS G2 SBL \footnotemark
\footnotetext{\url{https://store.waterlinked.com/product/underwater-gps-g2/}}. USBL can be installed on robotic system such autonomous underwater vehicle (AUV) \cite{dodge_turtlecam_2018} or autonomous surface vehicle (ASV) \cite{page_underwater_2021}. Dodge et al. \cite{dodge_turtlecam_2018} were able to follow a turtle with their AUV for several hours with an USBL. The drawbacks of the USBL systems are their prices, the transmitter size and the loss of accuracy in the shallow environment. 

Here, we focus on the accurate fine scale trajectory of our future target species, a juvenile green turtles evolving in the shallow reefs of Reunion Island. We chose the SBL Waterlinked UGPS G2 system to get the underwater positions. To overcome the range constraint of 100 m, we adapted the navigation system of the ASV to follow the acoustic transponder within this range. 

\subsection{Bathymetric survey}

Almost only professionals perform ASV bathymetry surveys, as they require expensive sensors such as an echosounder and a differential GPS for sub-centimetric positioning. The echosounder pings a signal to the bottom of the seafloor and measures the depth with the signal travel time and the signal velocity in water.
For instance, it can be used to map harbors or channels. For our ecological purposes, the bathymetric map can be compared with animal depth to understand water column use during specific behaviors  such as rest or feeding areas \cite{dodge_turtlecam_2018}. 

In the same way as other electronic systems, bathymetric sensors tend to be cheaper. For example, we first started with an ECT-400 echosounder \footnotemark \footnotetext{\url{https://www.echologger.com/products/single-frequency-echosounder-deep}} at \$3700 and we are now testing a S-500 by Cerulean\footnotetext{\url{https://ceruleansonar.com/products/sounder-s500}} at \$595. Several projects emerged in the past few years and offered ASV for bathymetry \cite{kimball_whoi_2014,raber_low-cost_2019,manda_low_2015,carlson_affordable_2019}. These projects are not easily reproducible. For instance, the Woods Hole Oceanographic Institution $Jetyak$ is not open source \cite{kimball_whoi_2014} . 
In Carlson et al. \cite{vlachos_software_2019}, the bathymetry accuracy is not specified but the depth map shows pixels around 10 x 10 m. In our application, we want to be able to discriminate small seabed components with at last 5 m radius. Price also limits the use of multibeam echosounders which still cost dozens of thousands of dollars. 

\subsection{Photogrammetric survey}

Photogrammetry enables the 3D reconstruction and mapping of a scene with overlapping images from different perspectives. Here, we propose an easy method for planning and validation of photogrammetric surveys performed with an ASV. For underwater purposes, archaeologists introduced it in 1968 \cite{drap_underwater_2012,pollio_applications_1968}. Recently, many research teams have applied underwater photogrammetry for scientific goals \cite{ferrari_3d_2017,marre_monitoring_2019,million_colony-level_2021}. Primarily focusing on small coral colonies with surveys made by divers, these studies give accurate coral surface estimation ranging between 2 to 19\% \cite{lavy_quick_2015}. Marre et al. \cite{marre_monitoring_2019} achieved an average model resolution of 3.4 mm. 

Lately, some studies have used ASV for photogrammetry surveys, but they need high computing resources and give less accurate resolution \cite{johnson-roberson_generation_2010}. Covering a larger area with an ASV is however made possible when images are coupled with accurate GPS position and orientation data. This additional information helps to run the model more quickly and accurately. Orthophotos can then be mapped on the bathymetry from the echosounder. 

Software improvement simplifies the computing process without the need of long and complex pre-processing with automatic camera-ordering or camera calibration. Several softwares are  available but to compare them is hard because it depends on the survey conditions and image quality \cite{vlachos_software_2019}. 

The drawbacks of using the ASV are the limited depth at which the bottom can be mapped, dependent on the light, image quality, turbidity, and condition at the sea surface. $Matisse$ \cite{arnaubec_optical_2015} is one of the rare photogrammetry software  available for underwater application. It is open-source and provides 3D and 2D models. 

This paper describes  the necessary tools to build and use an ASV with acoustic tracking ability as well as bathymetry and photogrammetry data collection. 

In the first section, we describe the ASV (mechanical, electrical and software parts). The validation and characterization Section presents each functionality of the ASV with field surveys. We provide the complementary information, mounting instruction, hardware, and software files as well as training datasets in a GitLab repository \footnotemark 
\footnotetext{\url{https://gitlab.ifremer.fr/sb07899/Plancha-ASV.git}}.



\section{ASV Description }

The ASV requirements are summarized in Table~\ref{tab1}. The hull is made from a paddleboard to be easily deployed, transportable and rugged. Depending on the deployment location, the needs and the different available resources, the ASV can be used with or without 3G/4G network. All functions are possible with or without internet although acoustic tracking is more complex without an internet connection and does not allow checking the tracking live. The global network system architecture of the ASV is detailed in Figure~\ref{fig_network}.

\begin{table*}
\begin{center}
\caption{ASV requirements for the various operations}
\label{tab1}
\begin{tabular}{| c | c |}
\hline
\textbf{Global}  &   \\\hline
\hline
  Handling & 2 people recommended  \\\hline
  Transport & $<$ 2.5 m (for aircraft regulation) \\\hline
  Deployment & From a small boat or the shore  \\\hline
  Environment  & Tested in tropical weather: Temp : 10 – 35°C  \\\hline
  Stance  & Stable for wave : 0.5m  /  wind : 20 kt \\\hline
  Guidance  & Autopilot and manual control  \\\hline
  Buoyancy & Can support $>$10 kg  \\\hline
  Communication  & Telemetry range $>$ 1 km  \\\hline
  Power limitation  & Motor under 2.5 kW  \\\hline
  \hline
\textbf{Surveys}  &   \\\hline
\hline
  Lifetime per survey & $>$2 h  \\\hline
  Speed & Between 0.5 and 1.2 m/s  \\\hline
  Navigation  & Autopilot allow following 1m transect 
 \\\hline
  Bathymetry sensor  & Single beam echo-sounder  \\\hline
  Photogrammetry sensor & Camera (e.g. GoPro) \\\hline
  Communication mode  & Cellular \& telemetry\\\hline
  \hline
\textbf{Tracking}  &   \\\hline
\hline
  Lifetime per survey  & $>$5 h  \\\hline
  Speed & about 0.8 m/s  \\\hline
  Mechanic   & 2 m between each hydrophone  \\\hline
  Sensor 1 & Acoustic geolocation system (SBL)  \\\hline
  Sensor 2 & Camera for behavior analysis \\\hline
  Communication mode  & Celular \& telemetry \\\hline
\hline 
\end{tabular}
\end{center}
\end{table*}

To be affordable and reproducible, all the electronical parts (excepted the echosounder) are commonly used components of robotics hobbyists and are easy to buy from general robotics sellers.

We divided this section into mechanical and electronic parts. In Table~\ref{tab:2}, we presented the main components with the total price of each ASV configuration. A complete BOM \footnotemark
\footnotetext{BOM link : \url{https://gitlab.ifremer.fr/sb07899/Plancha-ASV/-/blob/main/Documents/4_BOM.xlsx}} is provided. The mounting tutorials, wiring, CAO files, and installation configuration of the different software components are available on GitLab repository \footnotemark.
\footnotetext{Git link : \url{https://gitlab.ifremer.fr/sb07899/Plancha-ASV.git}}


\subsection{Mechanical part}

The main mechanical parts are a paddleboard, a waterproof case for the electronics, and a thruster support underneath the board. Some of the custom parts are made using a 3D printer. For the acoustic mode, arms are added to hold and immerse 4 hydrophones.

\begin{figure*}[htp]
\centering
  \begin{subfigure}{\textwidth}
  \includegraphics[width=\textwidth]{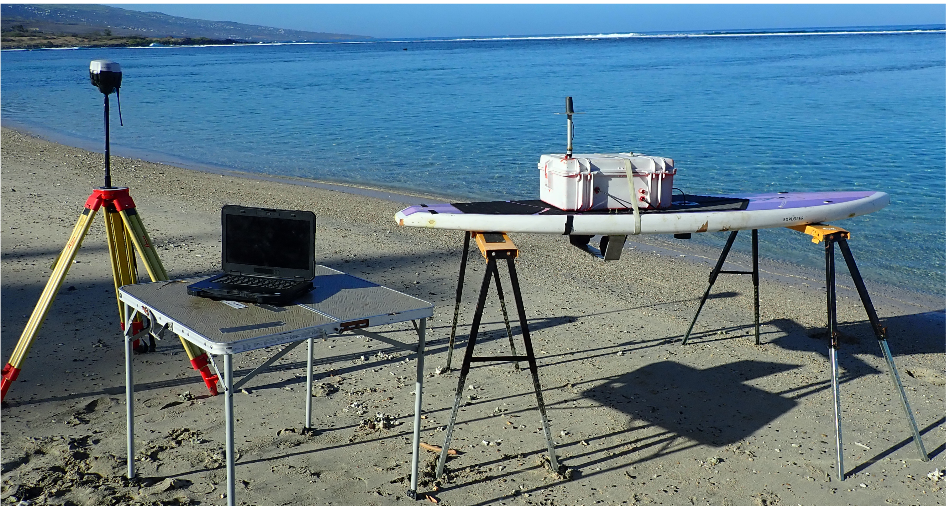}%
  \caption{ASV preparation for a survey in remote mode with the mobile GPS RTK base station (on the yellow tripod)}
  \end{subfigure}%

  \begin{subfigure}{\textwidth}
  \includegraphics[width=\textwidth]{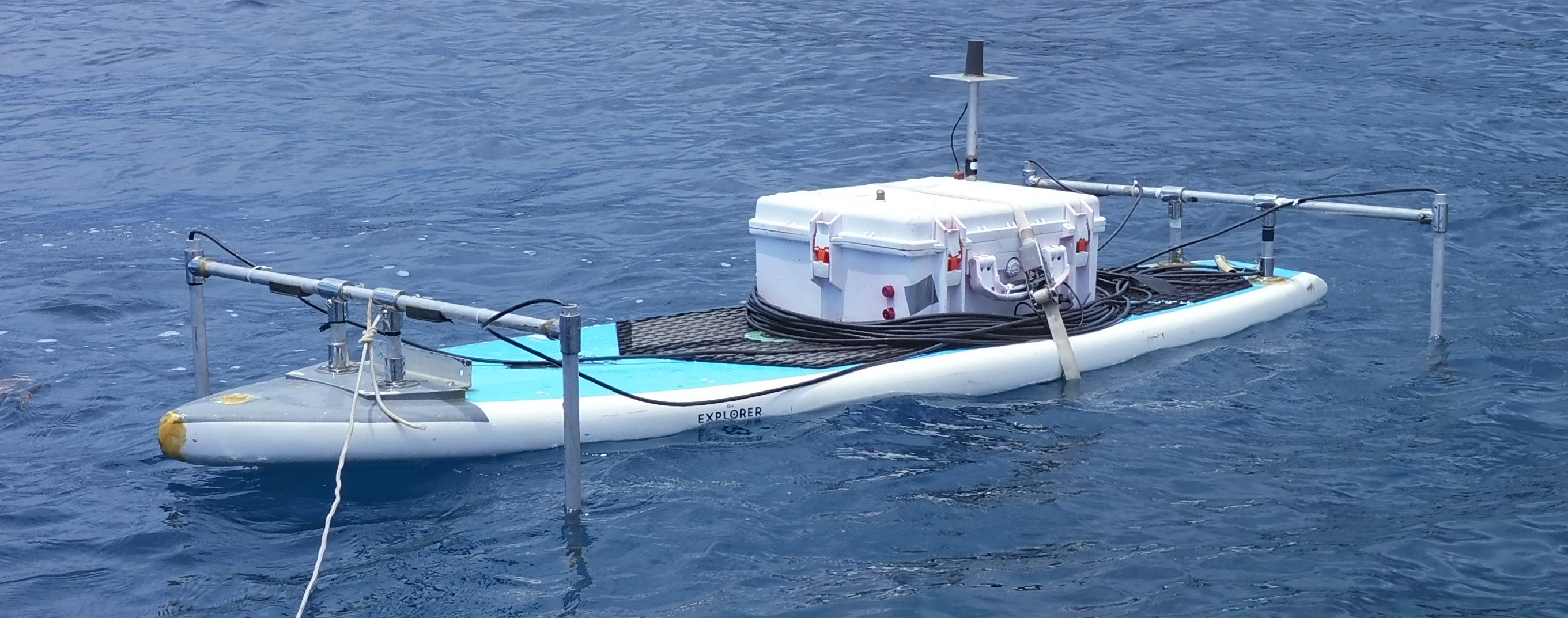}
  \caption{ASV in acoustic mode}
  \end{subfigure}%
  \caption{ ASV photos for the different modes: (a) Survey mode
for bathymetric and photogrammetric data collection and (b) acoustic mode for animal tracking with the four arms equiped with hydrophones.}
\label{photo_asv}
\end{figure*}


\begin{table*}
\setlength{\tabcolsep}{3pt}
\caption{Description of the main ASV parts for the different configurations and operations}\label{tab:1}
\centering
\subfloat[Different parts classified by mode and operation]{\label{tab:2}
\begin{tabular}{| c | c | c | c | c |}
\hline
\textbf{Global Mode} & \textbf{Component}  &  \textbf{Name} & \textbf{Nb} & \textbf{Unit Price} \\\hline
\hline
Electrical  & Fligth controller  &  Pixhawk cube 2.1 black & 1  &  \$315  \\\hline
  & GNSS RTK & Emlid reach M2 & 1 & \$499   \\\hline
  & Telemetry & RFD900 & 1 &  \$277   \\\hline
  & Radio command & RadioLink AT9S & 1 & \$129.99  \\\hline
  & Thruster & Blue robotic T200 & 2 &  \$179   \\\hline
  & ESC & Blue robotic Basic ESC  & 2 &  \$27   \\\hline
  & Battery & Tattu 14.8V 25C 4S 10000mAh & 2 &  \$149   \\\hline
Remote & GNSS RTK Base & Elmid reach RS2  & 1 & \$2199 \\\hline
  & GNSS radio communication  & Reach LoRa radio & 1 & \$118  \\\hline
Internet & 4G dongle & Huawei E3372 & 1 & \$50   \\\hline
         & Companion board & Raspberry pi 3B+ & 1 &  \$38   \\\hline
  \hline
Mechanical  & Hull & Paddle board 8", 80L & 1 &  \$250  \\\hline
& Waterproof case  & HRDR waterproof case  & 1 &  \$225.20 \\\hline
& Thruster support  & Custom aluminum support  & 1 &  \$150   \\\hline
& Cobalt Series Connector & Blue trail engineering Connector  & 2 &  \$67   \\\hline
\hline
\textbf{Surveys Mode}  &  &  &  &   \\\hline
\hline
Electrical  & Echosounder  &  ETC400 & 1  & \$3850 \\\hline
& Camera  &  GoPro Hero 7  & 1 &  \$349     \\\hline
\hline
Mechanical  & Echosounder holder & Printed custom part  & 1 &  ~\$20   \\\hline
\hline
\textbf{Tracking Mode }  &   &   &   &    \\\hline
\hline
Electrical  & SBL acoustic receiver system  &  Waterlinked Underwater GPS  &  1 &  \$2200  \\\hline
 & Acoustic beacon  &  Waterlinked locator U1  & 1  &  \$1500   \\\hline
 & Additional battery & Tattu 14.8V 25C 4S 10000mAh & 2 &  \$149   \\\hline
\hline
Mechanical  & Aluminum holding arm  & Aluminum tubs &  2  & \$200 \\\hline
\end{tabular}}

\subfloat[Price estimation of the ASV for the different modes. Only indicative, it does not include cheap components and spare parts]{\label{tab:3}
\begin{tabular}{| c | c | c | c | c || c |}
\hline
  & Global (G)  &  G + Surveys  &  G + Tracking  & G + Surveys + Tracking   & Remote \\\hline
  \hline
\textbf{Total} & $\sim\$2434$ & $\sim\$6634$  & $\sim\$6802$ & $\sim\$8672$ & add $\$2400$  \\\hline
\end{tabular}}
\end{table*}

\subsubsection{Hull, cases and thruster}

The hull is made with a simple paddle board of 8” and 80 L. Two thrusters are used and mounted on the protection support when the board is on the ground or in very shallow waters. This support is made in 5 mm marine aluminum to be robust and is screwed to the board. We place a support base screw and bolted it on both sides to be waterproof and robust. 
Cables are passed through the board thanks to two printed and coated cable entries. The echosounder support is also printed and potted in a hole drilled in the board. 

Electronic parts and sensors are in a waterproof case of $54 \times 42\times 22$ cm. The GPS antenna mast is made of aluminum and acts as a ground plane for the antenna. The echosounder is wired with the Binder 770 Bulkhead Connector and the plug from \textit{Blue Robotics}. For the wiring of the thrusters with high electrical current, we chose cobalt series bulkhead connector and the plug from \textit{Blue Trail Engineering}. \\

\subsubsection{Integration of acoustic part}

In our case, 4 hydrophones are needed for the acoustic system. We mount them with 2 aluminum holding arms separated by 2~m following the manufacturer recommendation (see Figure\ref{photo_asv}.b).  The first arm in the back of the board is composed of 5 aluminum tubes: 2 small tubes of 10 cm, 2 of 60 cm and 1 of 2 m. Connection between the 60 cm and 2 m tubes are made with stainless-steel elbow from marine hardware stores. Fixation of the arm and the board are made with stainless steel bases (on the board) and stainless steel Ts for the long tube. Bases are screwed and inserted into the board. As the space between the bases is smaller on the front, we reinforce the fixation by fixing the 2 bases on printed support which is potted on the board. To connect the 4 acoustic receivers, we used binder 770 bulkhead connectors and plugs from \textit{Blue Robotics}. They are already mounted on the acoustic electrical. \\


\subsection{Electrical part}

\begin{figure*}[ht]
\centering
\includegraphics[width=\textwidth]{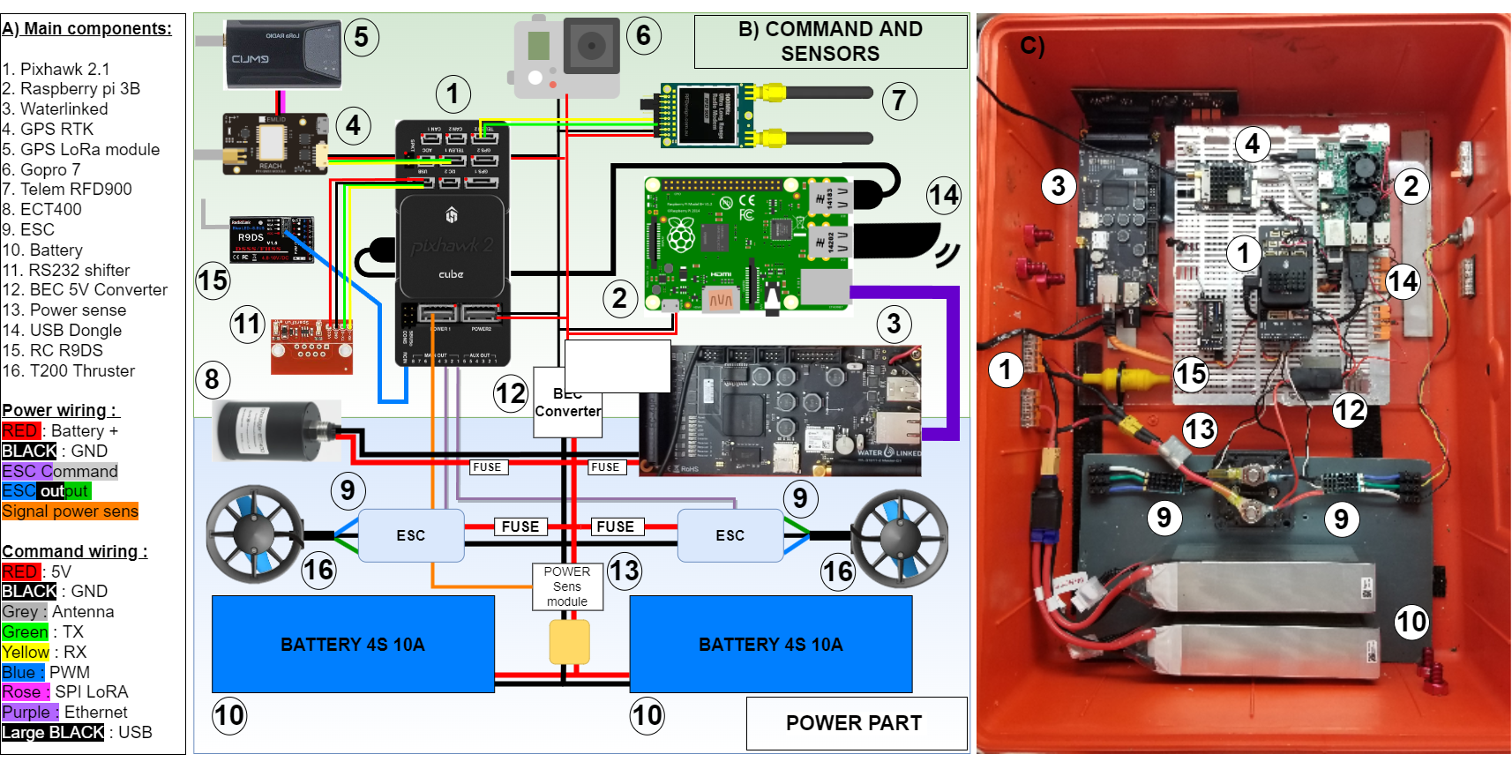}
\caption{ASV high level electrical diagram and electrical circuit. On the left (A), the corresponding numbers and names of the main parts. The colored names correspond to different wires on the electrical diagram. In the middle (B), the high level electrical diagram with main components and wiring. On the right (C), the electrical circuit with the corresponding numbers. Some components are fixed on the top of the case or outside and thus are not visible on this photo.}
\label{fig_elec}
\end{figure*}

For the electrical and software sections, we first described the power part and then the main components and sensors. In Figure~\ref{fig_elec}B), the power is represented by a blue background and the command and sensors by a green one. The core of this part is common for an ASV or a rover. It is composed of an autopilot (component 1), a GPS module (component 4) and communication systems (component 7). The entire electrical part, external sensors (Camera, echosounder) and the Electronic Speed Controller (ESC) for the thrusters are placed into a waterproof case (Figure~\ref{fig_elec}.C). Figure~\ref{fig_elec}A) represents the high level electrical diagram and a picture of the ASV electrical circuit with the annotated corresponding components.

%

\begin{figure*}[ht]
\centering
\includegraphics[width=\textwidth]{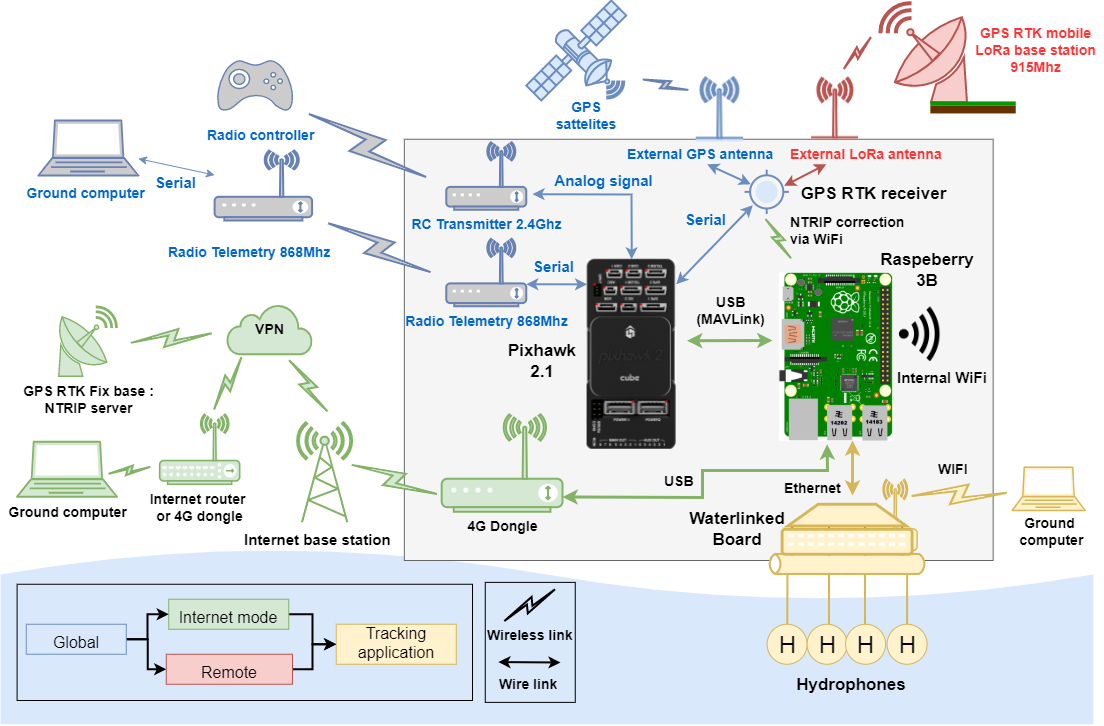}
\caption{Network diagram of the ASV showing how the autopilot gets and interacts the difference sources of information to perform the navigation of the ASV }
\label{fig_network}
\end{figure*}
%

\subsubsection{Power part}
The power part is composed of a minimum of two 4S/10Ah batteries (component 10 - Figure~\ref{fig_elec}), 2 electronic speed controllers (ESC) (component 12), 2 thrusters (component 16), 1 voltage monitor (component 13), 1 voltage regulator (component 12) and some fuses. Except batteries, all the components are from \textit{Blue Robotics}. 

The following section describes the software part and how the different components communicate with each other. A graphical summary is available in Figure~\ref{fig_network}.
\\

\subsubsection{Autopilot}

Autopilot or flight controller is the \textit{Pixhawk} 2.1 cube black (component 1 - Figure~\ref{fig_elec}). Except the camera and SBL, all the components and sensors are connected to the flight controller. The flight controller is powered through the 5V output of the voltage regulator. The power sense module provides information on battery voltage and electrical consumption. It is also connected to the \textit{Pixhawk}. 
The flight controller is configured with the open-source autopilot \textit{Ardupilot} rover V3.5 in "boat" mode. It handles the navigation rules and the configuration of hardware and sensors. The parameters of our configuration are given in the parameter file available in the GitLab repository \footnotemark. \footnotetext{Parameter file path: \textit{\path{https://gitlab.ifremer.fr/sb07899/Plancha-ASV/-/blob/main/Sotfware/Parameters/param_110122.param}}} 

\begin{figure}[htb]
\centering
\includegraphics[width=0.5\textwidth]{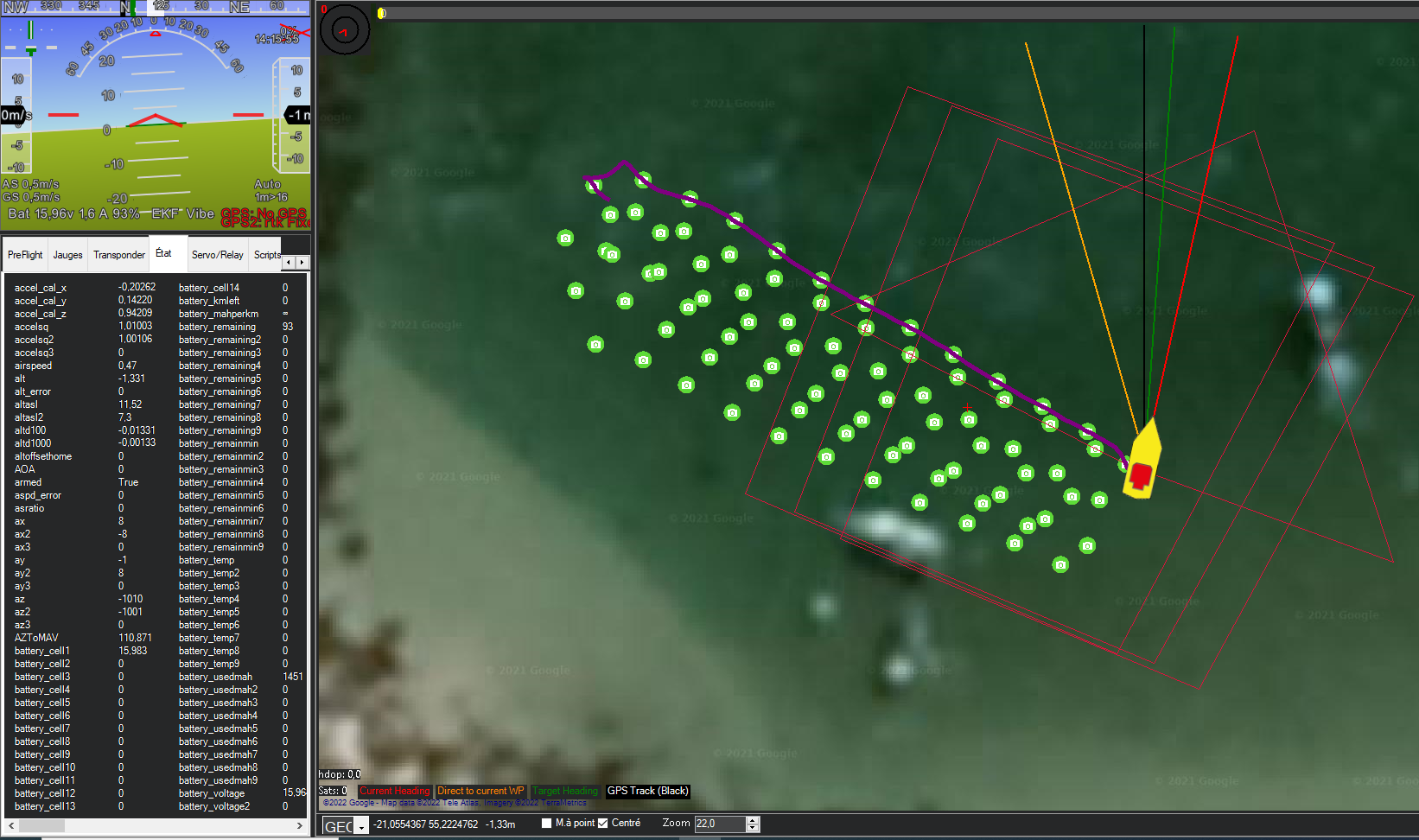}
\caption{Screenshot of Mission Planner during a navigation test in Saint-Gilles les Bains (Reunion island). The yellow boat shape corresponds to the ASV position. Purple line is its actual track and the green dots are positions where an external signal is sent to control a camera. }
\label{mission_planner}
\end{figure}

These settings depends on the board and the hardware used and a calibration should be done. 
The autopilot uses the mavlink protocol to communicate via USB to the companion computer and with radio telemetry to the ground-based computer. Ground control station software (GCS) is required to communicate with and control the autopilot. Different GCS are available and we used \textit{Mission planner}. GCS displays real-time variables and positions of the ASV. Mission Planner allows mission planning for the surveys and setting all the parameters of the vehicle (Figure~\ref{mission_planner}). More information on how to install and use it are available on the \textit{Ardupilot} website \footnotemark. 

\footnotetext{\url{https://ardupilot.org/copter/docs/common-choosing-a-ground-station.html}} A general tutorial about \textit{Ardupilot} rover is available on this link \footnotemark.\\

\footnotetext{\url{https://ardupilot.org/rover/docs/rover-first-drive.html}}

\subsubsection{Companion computer} 

The companion computer is a \textit{Raspberry Pi 3B} (component 2 - Figure~\ref{fig_elec}). It is powered by a 5V voltage regulator. The companion computer and the flight controller are connected with a USB cable for serial communication. The \textit{Raspberry Pi} has multiple roles: it communicates with the acoustic module and the flight controller and  allows running custom scripts used for sensors and ASV components.
During tracking mode, we run the \textit{Python} acoustic tracking script on the \textit{Raspberry Pi} which uses information from the flight controller and the acoustic modem. 
In internet mode, the connection is made using a USB 4G dongle. The companion computer then acts as a WiFi access point to share its connection. We set up and used a VPN with OpenVPN to be able to access the \textit{Raspberry Pi} with SSH via the internet. For more information on the \textit{Raspberry Pi} used as companion computer are available\footnotemark \footnotetext{\url{https://ardupilot.org/dev/docs/raspberry-pi-via-mavlink.html}}. Detailed information and procedure to install the \textit{Raspberry Pi} image are available on the GitHub repository \footnotemark. \\
\footnotetext{Software instructions link :  \url{https://gitlab.ifremer.fr/sb07899/Plancha-ASV/-/blob/main/Documents/2_software_insctructions.docx}}

\subsubsection{RTK GNSS}

We use a \textit{Emlid Reach} M2 \footnotemark 
\footnotetext{(\url{https://store.emlid.com/product/reachm2/}}
as a differential GNSS (component 4) with the possibility of Real Time Kinematics (RTK) (Figure~\ref{fig_elec}). Connection is made through serial communication with a telemetry port of the flight controller.
We power the $Reach M2$ with the micro USB connector connected to a 5V voltage regulator. A WebGui or a smartphone app is available to configure the $Reach M2$. 
In internet mode, the GNSS is connected to the WiFi access point of the companion computer and corrections are fetched through an online NTRIP server (for instance using a docker available here \footnotemark \footnotetext{(\url{https://github.com/goblimey/ntripcaster})}).
For remote mode, corrections are fetched using a LoRa link. In that case, a second GNSS receiver is set as a reference base and sends RTK corrections to the embedded GNSS. For that purpose, we used an \textit{Emlid} RS2 at a known position. The global setup of the GNSS module is available on emlid documentation\footnotemark.\\
\footnotetext{ https://docs.emlid.com/reach/reachview-3/connecting-to-reach} 

\subsubsection{Communication}

Different methods of communication are possible. For telemetry, we used a radio or internet connection. Even for the internet mode, we used radio telemetry as backup because this system is trustworthy. The Radio telemetry (component 7 - Figure~\ref{fig_elec}) allows for communication with the autopilot through ground station software via mavlink protocol. We chose the RFD900x module at 868 MHz which has a range of 20 km. It ensures a reliable link with the ASV and it is used in both modes. 

To control the board in manual mode and do some simple tasks such as arming/disarming the thrusters, we used an RC command using radio communication (RC model R9DS with radiofrequency at 2.4 GHz). The RC receiver is connected to the RCIN port of the flight controller. The RC radio command (component 15 - Figure~\ref{fig_elec}) is used to arm, disarm, and change mode. It is also used as a backup in case the other transmission systems fail.\\


\subsection{Additional Sensors} 
\subsubsection{Echosounder} 

The echosounder is the ECT400 by \textit{Echologger} \footnotemark (component (8)). It is a single beam frequency echosounder allowing bathymetry survey up to 50 m with a 5° beam. Its ground and power wires are connected to the output of the battery since its allowed power voltage spans from 8 to 75 VDC and thus does not requires any voltage regulation.
\footnotetext{(\url{https://www.echologger.com/products/single-frequency-echosounder-deep})}
The echosounder communicates by serial link with the flight controller. A RS232 level shifter is used to convert the output of the echosounder to a 5 V serial signal. Depth is stored in the ardupilot log as "DPTH" variable. It needs to be configured as described in the \textit{Ardupilot} turorial \footnotemark. \footnotetext{\url{https://ardupilot.org/copter/docs/common-echologger-ect400.html : Configuring the sensor}} \\

\subsubsection{SBL acoustic positioning} The SBL system is the underwater GPS G2 from \textit{Waterlinked} R100 (component (3)). It is composed of 4 acoustic receivers, a master board, and an acoustic beacon. The electrical board is connected to the \textit{Raspberry Pi} using an Ethernet cable. The input voltage range is between 10 and 30 V. We connected the board directly to the battery voltage by adding a 3 A fuse. The acoustic transmitter is the locator U1 \footnotemark. It works after manual activation and has 10 h lifetime. The SBL system has a 100 m of range in the standard version. The accuracy of the position given by the constructor is 1\% of the range, i.e., 1 m for this application.
A WebGui is available to configure the underwater GPS. The acoustic receiver array needs a specific baseline configuration.

\footnotetext{\url{https://store.waterlinked.com/product/locator-u1/})}
\textit{Waterlinked} recommends a distance of 2 m between each receiver. Distances between the acoustic receivers are measured on the paddleboard and set in the baseline configuration tab using the WebGui. For our application, orientation and position are fetched from the flight controller and sent by the companion computer. The settings "tab/top-side", GPS and compass have to be switched to \textit{External}.
To record the tracking, we used a custom \textit{Python} script run from the companion computer. The software and system integration information are explained in the documentation\footnotetext{\url{https://waterlinked.github.io/underwater-gps/quickstart/}}.
For our specific application, the procedure details are available in documentation folder \footnotemark. \footnotetext{Documentation folder :  \url{https://gitlab.ifremer.fr/sb07899/Plancha-ASV/-/blob/main/Documents/2_software_insctructions.docx}}

Position of the acoustic transmitter to the ASV is calculated with a signal Time of Arrival (TOA) algorithm between each different receiver. Then, the system needs the GPS position and heading of the ASV to calculate the geolocated position. To keep the acoustic transponder within the 100 m range, its position is defined as a new way point to be reached by the ASV. \\

\subsubsection{Camera} 

We used the \textit{GoPro} 7 black edition (component (6)). The camera is powered by 5V from the voltage regulator. Both photogrammetry and tracking modes rely on \textit{GoPro} 7 images. For the photogrammetry the \textit{GoPro} 7 faces down, whereas in tracking mode, it has a 30° angle from the vertical position. During the photogrammetric survey, the field of view of the \textit{GoPro} 7 needs to be as linear as possible. We set the ISO parameter to the lowest value (ISO 100) and the shutter speed at a high value to get clear images and the \textit{GoPro} 7 is set in video mode. A minimum of 70\% of coverage is required between two pictures for photogrammetry. To set the space between transects, we used an excel file \footnotemark 
\footnotetext{\url{https://github.com/pierregoge/Plancha-ASV/blob/main/Sotfware/Photogrammetry/Spacing_between_transect_calculator.xlsx}}
calculating this space as a function of the depth of the survey area and the coverage needed. The distance between transects will also highly depend on the navigation accuracy capabilities. 
More information on the photogrammetric mission planning and pre-processing are available in the "prototype and survey results" Section and on the Github repository \footnotemark.


\section{Prototype validation and survey results}

To illustrate the potential applications of the ASV, we present some survey results. The validation of the ASV (e.g. accuracy of the trajectory) and the power consumption estimates are provided as Supplementary Materials.
All the data and software presented in the section are fully available here\footnotemark
\footnotetext{Illustration examples link : \url{https://gitlab.ifremer.fr/sb07899/Plancha-ASV/-/tree/main/Features_example}}.

\subsection{Autonomous acoustic tracking}

The acoustic tracking feature allows us to get a fine scale live trajectory and an active tracking of the underwater acoustic beacon (U1 Locator). For ethical reason, we did not test and deploy the U1 Locator on a marine turtle which is a protected species although it is the target species for our application. We deployed the beacon on a freediver who was asked to mimic turtle diving behavior. The survey is carried out at Cap Lahoussaye (-21.017348°N, 55.238212°E). The locator U1 was fixed on a diver's chest with a 50~cm offset from his body towards the seabed so the locator is still underwater when the diver is at the surface and to avoid any loss of the acoustic signal. It is noteworthy that even with this 50 cm offset, we denote more spikes or signal losses when the diver is at the surface. We set the navigation rules to update the distance between the ASV and the diver every second and lower than 5 m.  \\

The acoustic tracking feature propose here can be used for other applications such as tracking AUV or any other animals with a limited swimming speed. The next subsections present the tracking procedure, the data processing, and the results of the survey example. 

\begin{figure}[h]
\centering
\includegraphics[width=0.5\textwidth]{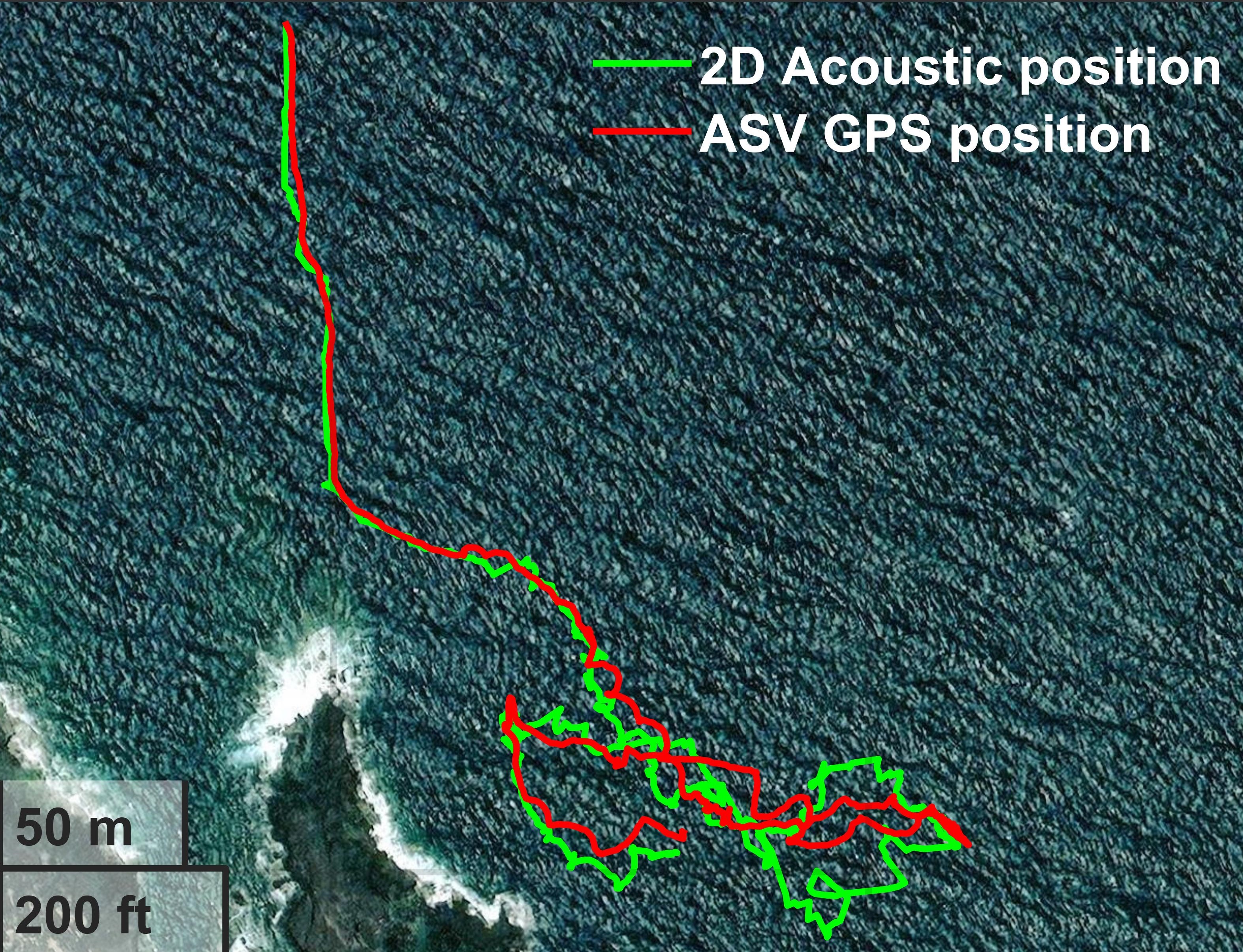}
\caption{ASV tracking of a freediver. Green track is the underwater acoustic position. Red is the ASV position}
\label{fig_tracking}
\end{figure}

\subsubsection{Protocol}

The \textit{WaterLinked} system does not save the trajectory and only displays it on their WebGui. In their github\footnotemark \footnotetext{\url{https://github.com/waterlinked/examples}}, \textit{WaterLinked} gives example scripts in \textit{Python} to use or save the data that can be run directly from a laptop. For tracking and logging, we developed our own logging scripts \footnotemark \footnotetext{\url{https://gitlab.ifremer.fr/sb07899/Plancha-ASV/-/tree/main/Sotfware/Tracking}}. 

The tracking algorithm \footnotemark enables the calculation of waypoints and their transfer to the autopilot. To start the tracking mode, the user needs to run the command on the \textit{Raspberry Pi} (see software instruction). \footnotetext{Tracking script in the raspberry. File name: \textit{main\_tracking.py}}
For the calculation of the next waypoint, the algorithm works as follows: Position and heading are read from the Flight controller of the ASV. It is then sent to the SBL module to calculate the position of the acoustic beacon. The Raspberry then sends a request for the position of the acoustic tag, compares the positions and decides if the ASV needs to move. If the acoustic beacon and the ASV are too close, the autopilot switches to hold mode and stands in its position. If the beacon moves away from the board and the threshold distance is exceeded (here 5 m), then a new position is sent to the autopilot which tries to reach it. Tracking parameters are stored in the Raspberry Pi\footnotemark
\footnotetext{Parameter file path in the \textit{Raspberry} : /idocean/parameter.json file}.

\subsubsection{Data processing}

Tracking data of the 3D position of ASV and acoustic beacon are logged in the Raspberry. A \textit{MATLAB}\textcopyright\ script was developed to analyze, filter, and plot the data.
We filter out the position data for which the standard deviation of the position estimates are larger than 3~m. A linear interpolation is performed to filter the positions of the acoustic track. 
\\

\footnotetext{Processing script in git: \url{https://gitlab.ifremer.fr/sb07899/Plancha-ASV/-/blob/main/Features_example/test_tracking_26_10_21/code/main_acoustic_tracking_20_10_21.m}}

\subsubsection{Results}

Figure \ref{fig_tracking} shows a 25-minute sequence over which the free-diver is successfully tracked in 3D. This example demonstrates the ability of this system (ASV + acoustic beacon) to collect precise underwater positions that can be used as reference data for animal tracking applications.

\begin{figure}[h]
\centering
\includegraphics[width=0.5\textwidth]{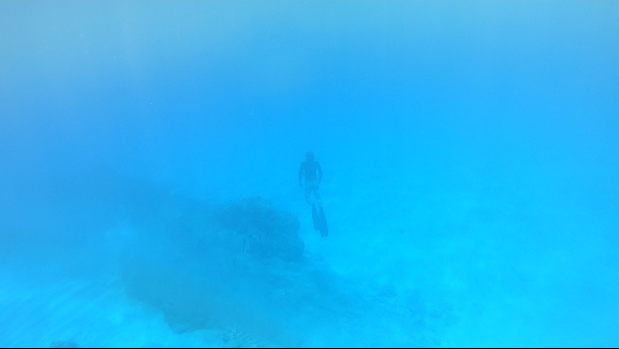}
\caption{Screenshot of the GoPro 7 footage during the tracking test when the diver is going up to the surface. As the seawater is turbid, it limits the ability to use the video for further behavioural analyses.}
\label{fig_gopro}
\end{figure}

For further video analyses, the image quality highly depends on the underwater visibility and the distance to the target. Figure \ref{fig_gopro} shows that videos can only be used when the visibility is good so it enables behavioral analyses. Moreover, when the ASV is close to the target, it stays in holding mode and drifts and it can lose the target of the camera field of view.

\subsection{Bathymetry survey}

Information extracted from bathymetric data depends on sensor specifications, but is also strongly related to the area topology and spacing between collected points.
Primary parameters such as the maximum measurement range, the sampling frequency or sensor errors have been fixed during the design phase.
For each survey, an \textit{a priori} knowledge of the sea ground topology is required to define the aimed data spatial distribution over the survey area.
Knowing the average depth and type of ground (e.g. large rocks, sand rift, corals, ...) will help to adjust the spacing between points. The spacing between strips has also to be adapted to the targeted map resolution.

Several standards define and classify the \textit{quality} of bathymetric surveys. For instance, in \cite{iho2020} (section 7.3, Table I), the \textit{International Hydrographic Organization} proposes five categories based on the overall accuracy, the area coverage, and the types of features that can be detected to help classify the \textit{quality} and \textit{goals} of a survey. We use these categories to define our specifications.

The next sections present the protocol, the processing stages, and the final results for a bathymetric survey with an illustration from a survey carried in 2020 on the north shore lagoon of Europa island in the Mozambique Channel.\\

\subsubsection{Protocol}

We set up the survey to meet the requirements cited in \cite{iho2020} and described in Supplementary Materials. This enables us to reach the \textit{order 1a} category, i.e.  data in  harbors, harbor approach channels, coastal areas or inland navigation channels, with a limitation to areas with less than 100 m water depth. 

The area of interest was a lagoon in Europa Island. Bathymetry in this area has been estimated using hypersectral and LiDAR data collected by the Litto3D Océan Indien project in 2019 \footnotemark
\footnotetext{data accessible here: \url{https://oceans-indien-austral.milieumarinfrance.fr/Acces-aux-Donnees/Catalogue\#/metadata/6b796349-d56e-44c3-b572-d5488250637e}}  (see section \ref{sec:result_bathy_comp}). From these data, the depth in the area of interest is ranging from 1 to 10 m. 

From these specifications and to reach the \textit{order 1a} bathymetry standard, the aimed survey area is a rectangle of 49 m $\times$ 115 m, with a center coordinates at -22.340984°N, 40.337634°E. The parameters to configure the ASV autopilot have been set as follow: 
\begin{itemize}
    \item 24 transects in the direction of the largest dimension (width), with a 2-m spacing.
    \item a target cruise speed of 1 m/s.
    \item a depth sampling rate of 2 Hz.
\end{itemize}
These result in a grid of $24 \times 228$ points over the survey zone, in which the \textit{bathymetric pixels} have a diameter ranging from 9 cm to 90 cm for depth ranging from 1 m to 10 m. Pixels have a widthwise spacing of 0.5 m and a lengthwise spacing of 2 m.\\

\subsubsection{Data processing}
The data are retrieved from the autopilot log file which includes all information, status and measurements done by the ASV during the survey.
A first step is to discard any unnecessary data to keep only the echo-sounder, GPS, and IMU data over the survey area.
To achieve an accurate depiction of the seabed, a pre-processing stage is required to correct and filter the measured depths. The raw data processing includes the following steps:
\begin{itemize}
    \item From the ASV attitude (roll, pitch, yaw) given by the IMU sensor, all points for which the pitch and roll angles are greater than a defined 10° threshold are removed.
    \item Using a sliding median-filter, depth values that are outside a certain range around the median depth value computed along the sliding window are removed.
    \item GPS data with position offsets between the GPS antenna and the location of the echo-sounder on the ASV are corrected for the 3 axis.
    \item Retrieve the true location of the measured depth on the sea floor by correcting the surface GPS positions with ASV attitude.
    \item Correct the recorded depth values with the ASV attitude, the local datum and the geoid of the survey zone, to eventually get a compensated and georeferenced depth map.
\end{itemize}

A minimal working example in $Python$ is available on the git repository\footnotemark\ associated with this article
\footnotetext{Example bathymetric data processing script in git : \url{https://gitlab.ifremer.fr/sb07899/Plancha-ASV/-/blob/main/Sotfware/Bathymetry/Compute_depth.py}}
\\
\subsubsection{Results and comparison with prior data }
\label{sec:result_bathy_comp}

\begin{figure}[h]
    \begin{subfigure}{0.5\textwidth}
    	\centering
    	\includegraphics[width=\linewidth]{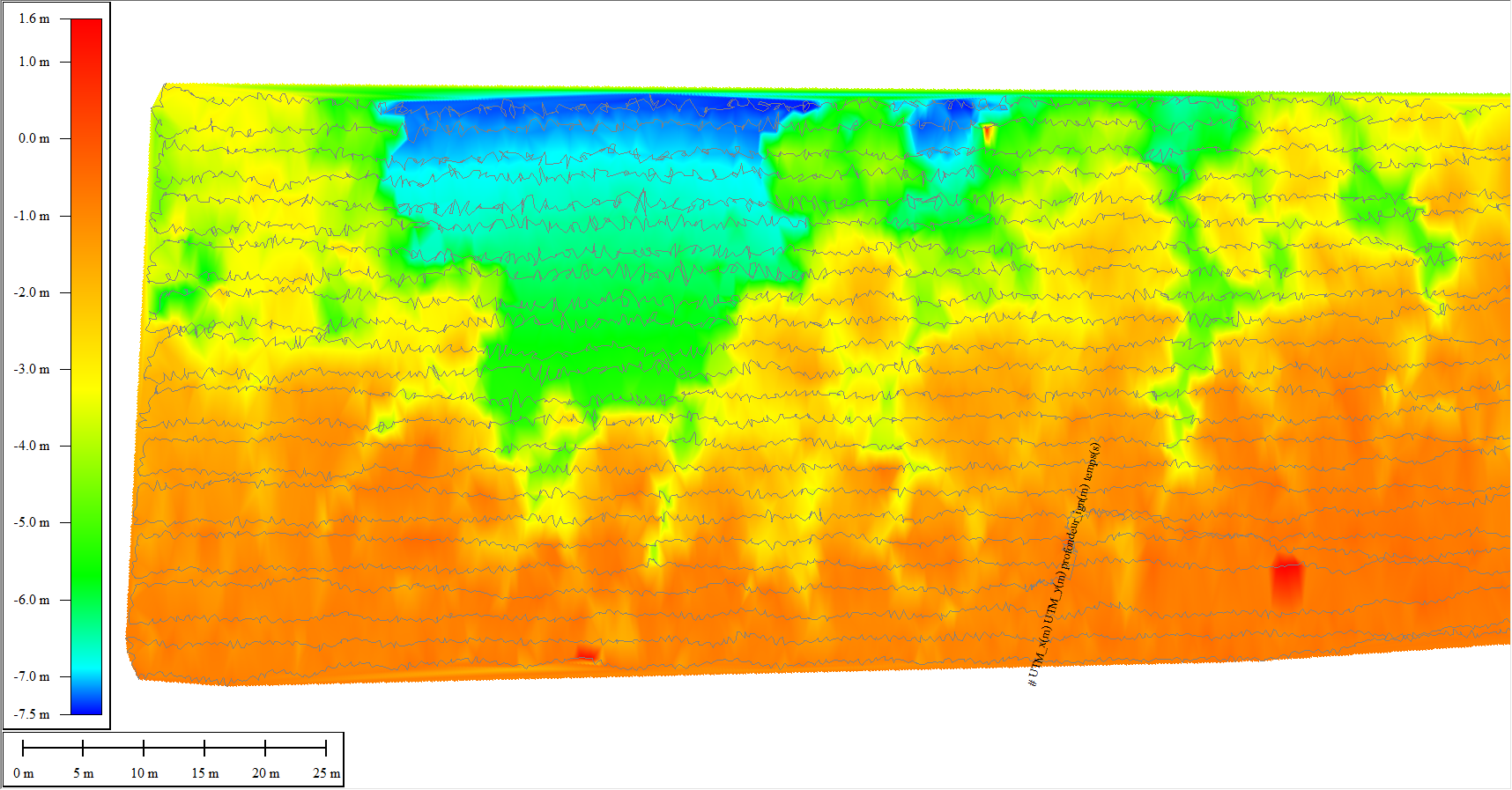}
    	\caption{Computed sea depth map with overlaid ASV paths (grey lines). Map generated with Global Mapper\textregistered }
    \end{subfigure}
    \hfill
    \begin{subfigure}{0.5\textwidth}
    	\centering
    	\includegraphics[width=\linewidth]{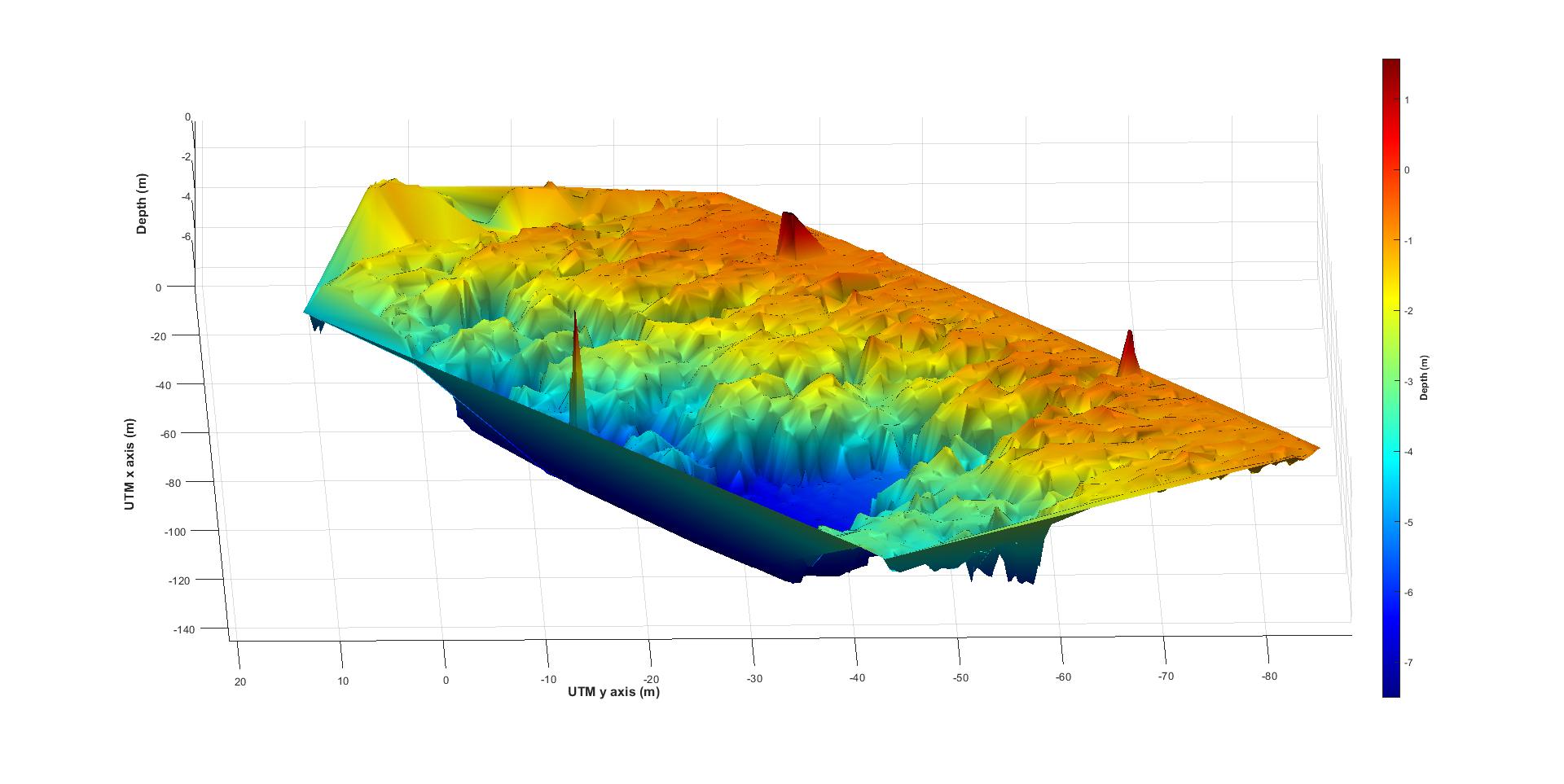}
    	\caption{Same bathmetric data with 3D-projection and Delaunay triangulation. Plot generated with \textit{MATLAB}}
    \end{subfigure}
    
    \caption{Bathymetry results of a survey done in 2020 in Europa Island with the ASV} %
    \label{fig_bathy}
\end{figure}

For the survey mentioned above, Figure~\ref{fig_bathy} shows different depth estimates of the same pre-processed data set. In Figure~\ref{fig_bathy}(a), the depth map has been automatically computed using the \textit{Global Mapper}\textregistered\ software. Overlaid gray lines represent the ASV path extracted from raw GPS data. Figure~\ref{fig_bathy}(b) is a 3D-projection of the same bathymetric data set build with \textit{MATLAB}\footnotemark.
\footnotetext{Example script in git : \url{https://gitlab.ifremer.fr/sb07899/Plancha-ASV/-/blob/main/Features_example/test_bathy_europa_09_10_20/code/main_plot_bathy_09_10_20.m}}

\begin{figure*}[h]
    \begin{subfigure}{0.3\textwidth}
    	\centering
    	\includegraphics[width=\linewidth]{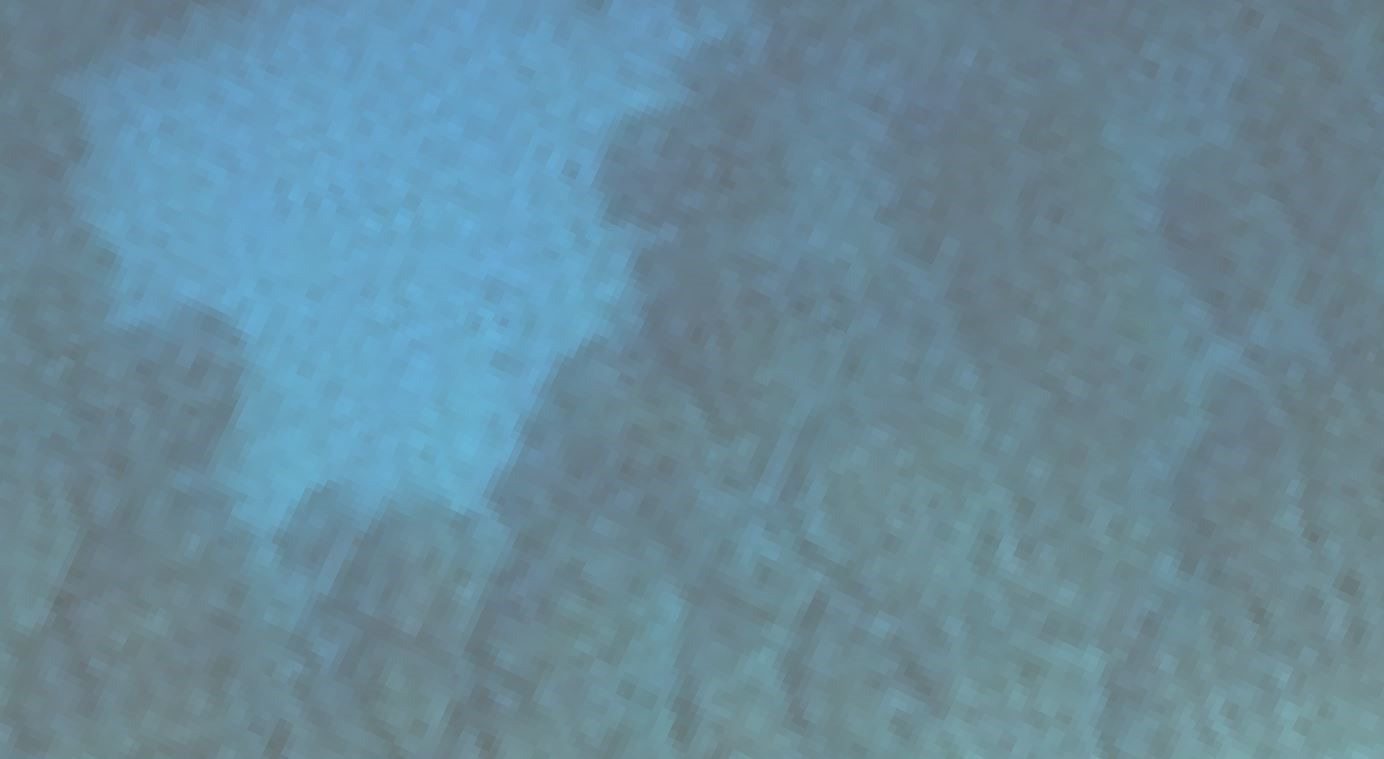}
    	\caption{Satellite imagery Pleiades © CNES 2019, distribution Airbus DS / Spotimage}
    \end{subfigure}
    \hfill
    \begin{subfigure}{0.3\textwidth}
    	\centering
    	\includegraphics[width=\linewidth]{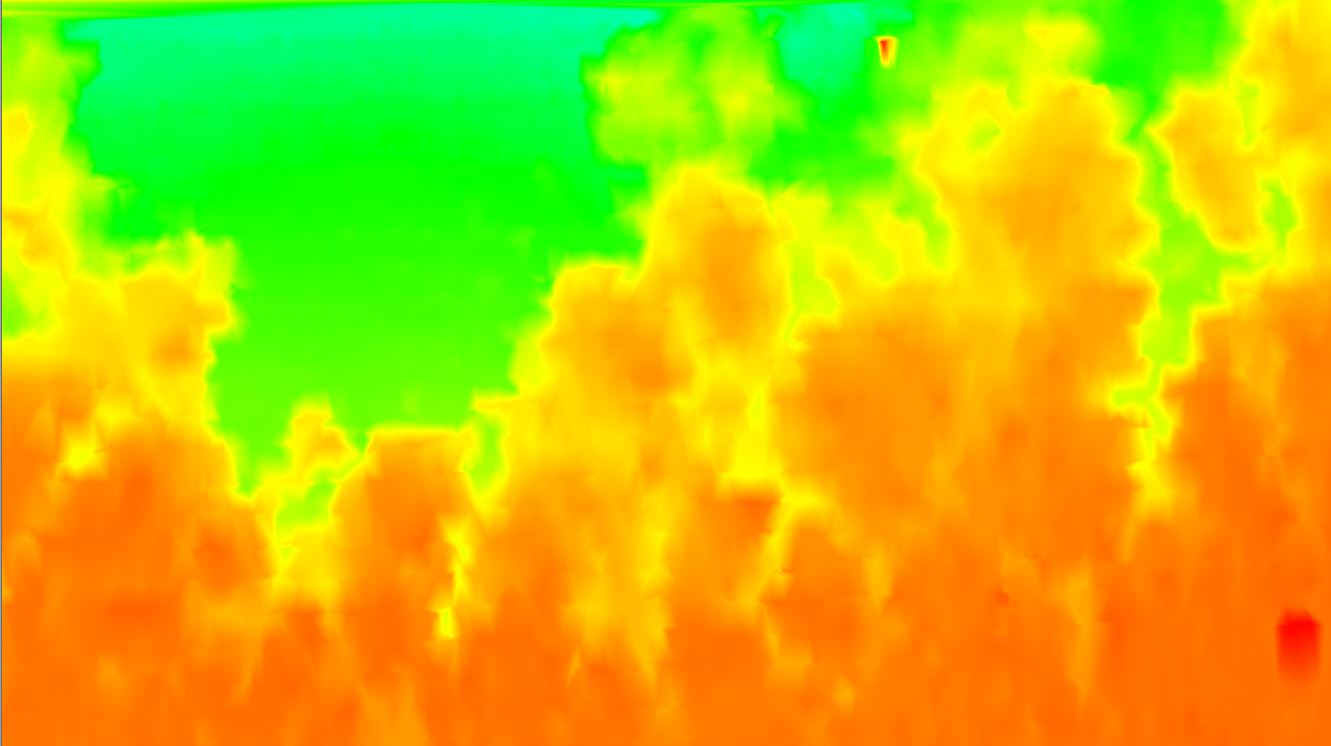}
    	\caption{ASV's bathymetry with the ECT400 echo-sounder (2020 campaign)}
    \end{subfigure}
    \hfill
    \begin{subfigure}{0.29\textwidth}
    	\centering
    	\includegraphics[width=\linewidth]{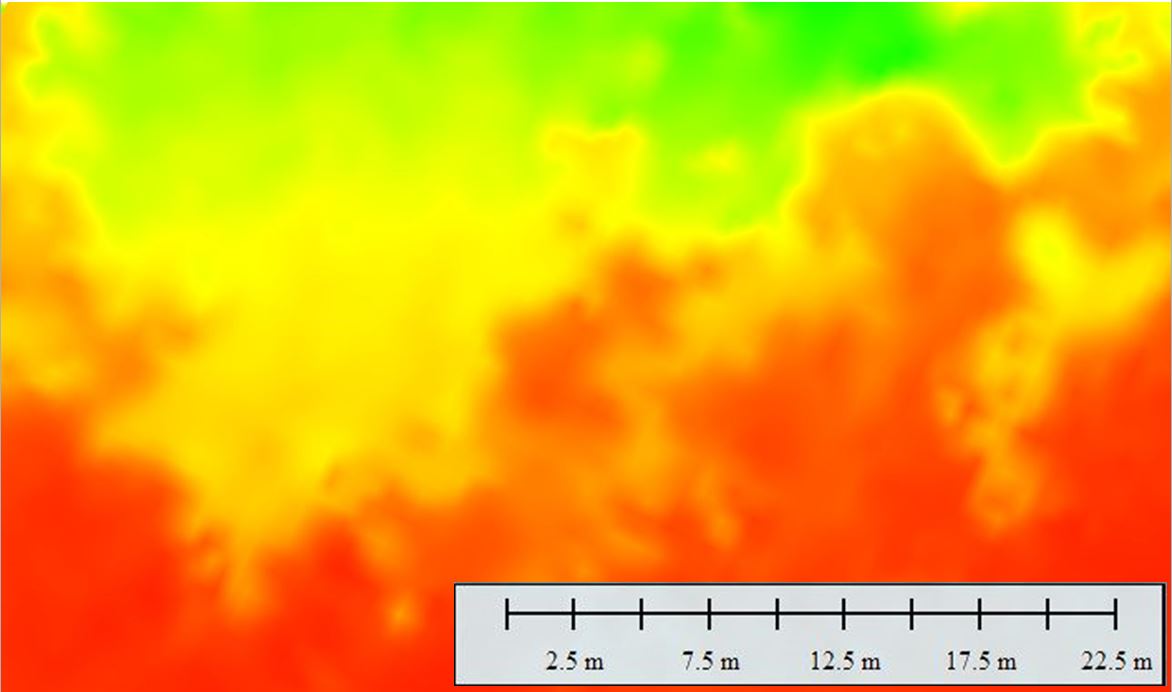}
    	\caption{LIDAR / Hyper-spectral bathymetry (2010 campaign)}
    \end{subfigure}
    
    \caption{Three different representations of the sea floor in the survey area located inside the Europa's lagoon to compare the results from the bathymetry estimated from the ASV data to the one estimated from hyperspectral and LiDAR data } %
    \label{fig_bathy_vs_sat}
\end{figure*}

To illustrate the benefits of using a single-beam echo-sounder on such ASV, we compare three different techniques that have been used to analyze the sea floor of the Europa lagoon (Figure~\ref{fig_bathy_vs_sat}).
Maps are drawn for a portion of the survey area discussed before. Figure~\ref{fig_bathy_vs_sat}(a) shows the satellite imagery of the surveyed area.
Figure~\ref{fig_bathy_vs_sat}(b) is a zoom on the map shown in Figure~\ref{fig_bathy}(a) representing the ASV bathymetry data. Figure~\ref{fig_bathy_vs_sat}(c) is the bathymetric data estimated from hyperspectral and LiDAR data collected in 2019 on the same area (Litto3D Océan Indien project).

A strict comparison of feature resolution and depth accuracy obtained with the three methods above is out-of-the-scope of this paper.
Such analysis would require special attention to the different geodesic reference frames used, the level of depth correction applied, whether it includes or not environmental/experimental parameters (i.e. temperature, salinity, the effect of tides, ...), and eventually to the interpolations errors introduced by the different spatial distribution of each data set.

However, a qualitative analysis is enough to confirm that the ASV bathymetry gives an accurate depiction of the seabed topology in this area as compared to the satellite imagery. We observe a similar bathymetry between the ASV data and the hyperspectral/LiDAR data but with a higher level of details for the ASV bathymetry.
Although aerial hyper-spectral techniques have the advantage of covering larger zones in a much shorter time, for smaller areas, deploying single-beam echo-sounders on such ASVs can be cheaper and a more practical solution with better resolution.
Finally, mounting this type of echo-sounder on an ASV instead of a boat has the advantage of much regular and dense sampling patterns, as well as the opportunity to investigate areas where it is too shallow for navigation.

%

\subsection{Photogrammetry survey}

Camera images collected over the survey area can be used to obtain photogrammetry data. Here we described the protocol, the data processing, and the results for this type of surveys.  \\

\subsubsection{Protocol} 

To obtain the best possible results for the photogrammetry reconstruction, it is required to calibrate the camera. Indeed, to prevent lens distortion, the parameters of the lens and image sensor of the GoPro camera have to be estimated. For this calibration, multiple images of a 9 by 7 square chessboard pattern are taken in different positions and with varying angles. Camera parameters are set to full resolution. The photogrammetry software, $Matisse$, has a built-in calibration process which proposes to compute and save the camera model. We can choose between different \texttt{Distortion models} in the camera calibration settings to correspond to the fisheye distortion of the GoPro. 
To obtain a three-dimensional reconstruction of the survey area, it is necessary that each image must have an overlap greater than 70\% with other images and photos are clear without surface effects on the seabed or ASV shadow.

Using the survey and camera information (i.e., field of view of the camera, sea depth), it is possible to define the distance between transects  that approximately satisfies the first condition. We propose a tool\footnotemark \footnotetext{\url{https://github.com/pierregoge/Plancha-ASV/blob/main/Sotfware/Photogrammetry/Spacing_between_transect_calculator.xlsx}} to estimate this distance. It does not take into account the sampling frequency of the camera and the speed of the ASV. In the example given in this paper, the speed of the vehicle is set to 0.8 m/s, the distance between transects is set to 2 m, and the depth in the studied area varies between 2 and 4 m. \\

\subsubsection{Data processing}

Underwater images suffer from various color degradation (correlated with the depth at which the image was taken, light fluctuations due to sunlight refraction etc). Matisse 3D carries out color and illumination corrections in a pre-processing mode. In our case, since the illumination was pretty uniform, we have checked only the \texttt{Correct colors for underwater attenuation} option while limiting the size of the images to 4 megapixels.

Once this preprocessing terminated, the reconstruction with Matisse 3D can be run with the post-processing mode. In order to obtain the higher reconstruction resolution, we choose the \texttt{3D Dense} algorithm.

Matisse offers the possibility to use the GPS positions and orientations of the photo metadata in order to improve the result of the photogrammetry process. This piece of information is available through the ASV log but we do not use this functionality yet which need one more pre-processing step to set the metadata of the images. \\

\subsubsection{Results}
\label{sec:result_photog}

\begin{figure*}  
\begin{subfigure}{\textwidth}
	\centering
	 \includegraphics[width=\textwidth]{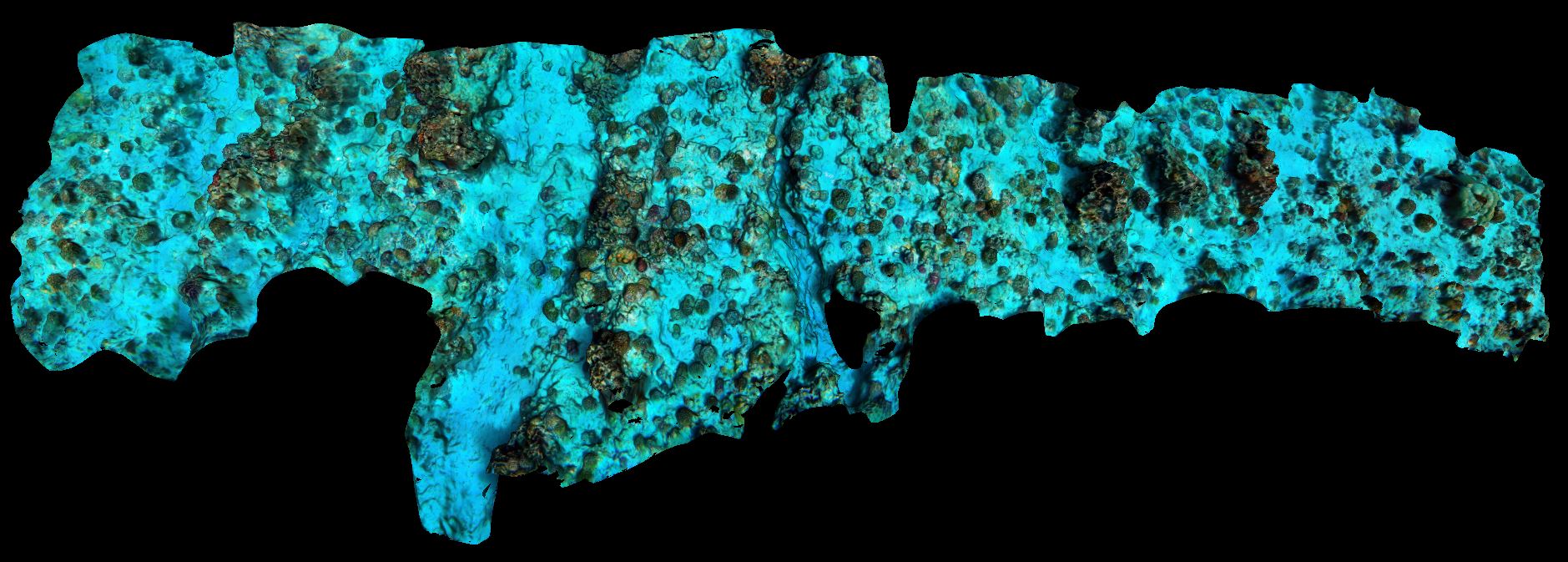}
	\caption{Top view reconstruction}
\end{subfigure}
\bigskip  
\begin{subfigure}{0.5\textwidth}
	\centering
	\includegraphics[width=\linewidth]{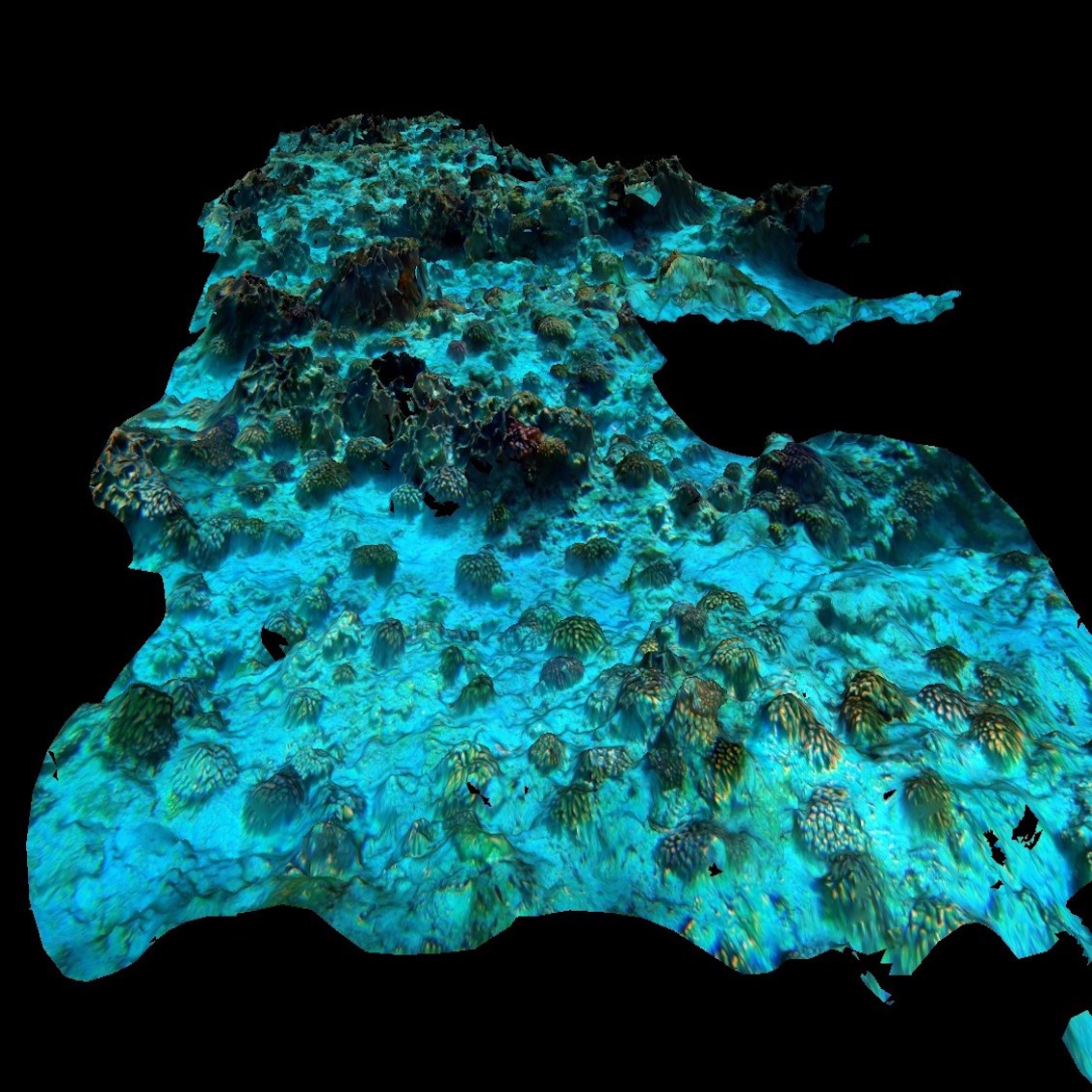}
	\caption{Side view reconstruction}
\end{subfigure}
\begin{subfigure}{0.5\textwidth}
	\centering
	\includegraphics[width=\linewidth]{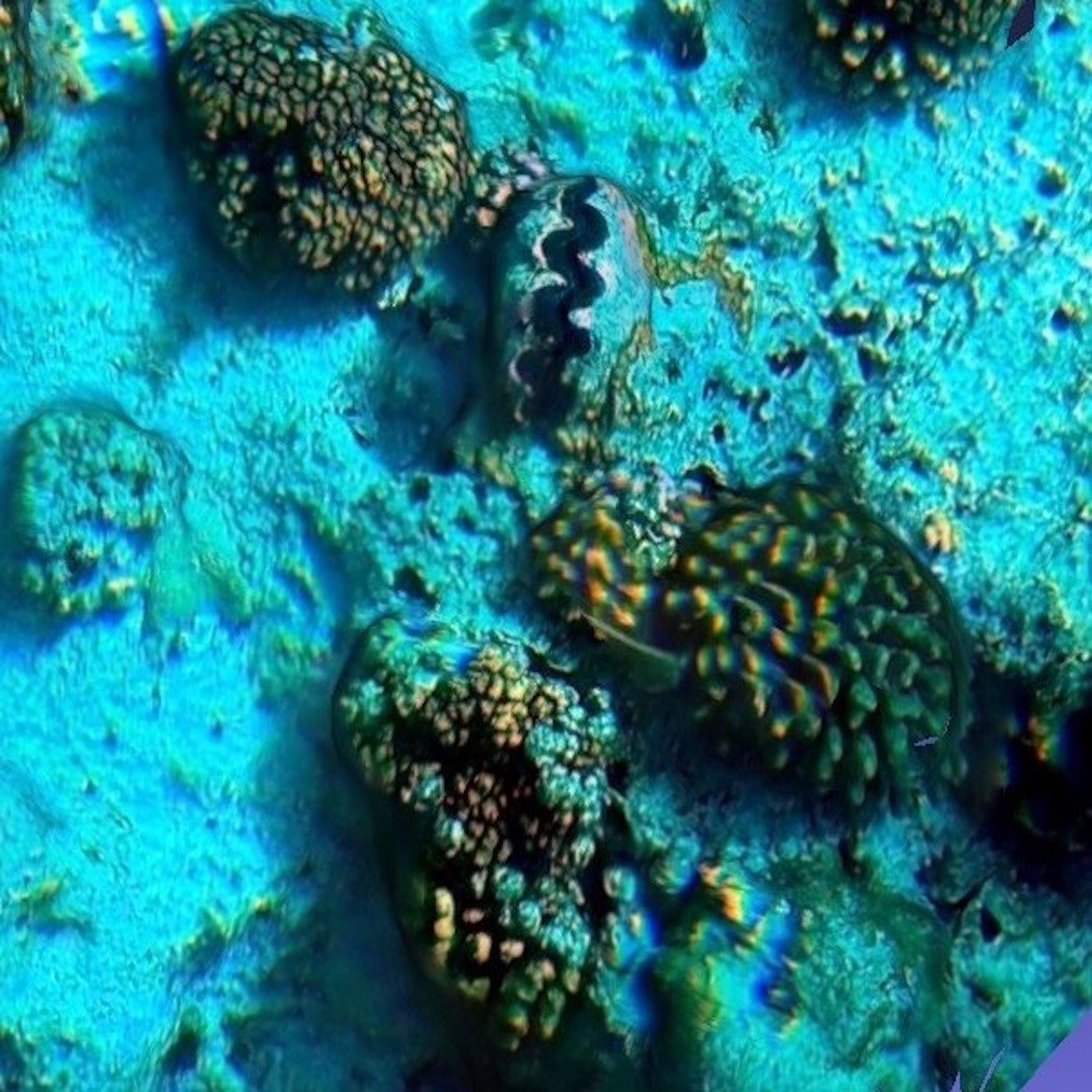}
	\caption{Zoom on a giant clam and on a $Pocillopora$ coral}
\end{subfigure}

\caption{Different views and zooms of the photogrammetry calculated from 70 images collected by the ASV during a field survey in Europa island in 2020.} 
\label{fig_photog_reconsctruction}
\end{figure*}

A result of a photogrammetry process on 70 images taken in Europa island is shown in Figure~\ref{fig_photog_reconsctruction}. Although the photos are all taken from the sea surface and the angle between the GoPro and the seabed remains unchanged (except for small variations due to the waves), the three-dimensional reconstruction can be performed and numerous elements characterizing the morphology of the study area can be identified (Figure~\ref{fig_photog_reconsctruction}). The geological faults are reconstructed, as well as numerous coral specimens of \textit{Acropore Massive}, \textit{digitised} and other elements such as a specimen of \textit{Clam}.

It must be emphasized that the 3D reconstruction and the level of details is strongly correlated to the amount, position, and angle at which the photos were taken, i.e., some portions are simply not reconstructed (black areas on the image) due to a lack of images or are degraded (black spots on the side of corals). 

\section{Conclusion}

This paper fully describes the hardware, software, and data processing tools for an autonomous surface vehicle. The ASV is able to perform:

\begin{itemize}
\item{an autonomous navigation with an autopilot}
\item{an autonomous acoustic tracking with an acoustic SBL system}
\item{a bathymetric survey with a single beam echosounder for depth $<$ 50m.}
\item{a photogrammetric survey with a low-cost camera } \\
\end{itemize} 

All the components and mechanical parts are chosen to be low-cost, easy to find, and easy to build. Regarding softwares, the firmware, flight controller, and in-house development are open-source. 

There are limitations to the ASV. For example, it is not designed to be used in rough sea and weather conditions. The ASV has flipped when deployed in windy conditions ($>$20 kt) and with choppy waves breaking ($\approx$0.3 m). 

In parallel to the description and the validation sections, we provide a Git repository containing all the documents, instructions, and files to reproduce this ASV. We illustrate the different features exposed for our applicaton with dedicated field surveys. The ASV can be deployed in different environmental conditions, with or without internet. The radio telemetry system allows to control and operate the ASV with a few kilometers range. For inhabited coastal regions such as Reunion island, the ASV never loses its internet connection within the survey area ($<$1 km from the coastline). The consumption of the ASV allows more than 4 h of survey time with two 4S batteries (10 Ah each). These batteries are compliant with air transportation regulations and makes the board easy to travel with a surf bag.

To summarize, Plancha ASV is reliable, easy to use, reproducible, and customizable. The system is small and light, and can operated by two operators which is advised to be able to recover the board in case of an issue. With telemetry and ground control software, the ASV can be followed in real-time during the survey with a laptop. This software also offers to create survey missions, change the parameters, and calibrate the ASV. The \textit{Ardupilot} flight controller logs the flight data and makes analyses easy with the appropriate tools.
 
With its high buoyancy and the space available on the board, other sensors, batteries, and other functionalities can be added. New software integration is straightforward thanks to the \textit{Raspberry pi} as a companion computer and \textit{Pixhawk} 2.1 with \textit{Ardupilot}.  

These functions and features prove that low-cost ASV can be used for environmental and ecological purposes and provide accurate monitoring. As far as we know, this is the first time that an ASV is used to track an acoustic beacon using a low-cost SBL system. This ASV can be used to provide accurate fine-scale trajectories of underwater animals even on shallow depth and to simultaneously collect environmental information such as bathymetry and photogrammetry.


\section*{Acknowledgement}

We are grateful to Steven Le Bars and Vincent Macaigne from ID OCEAN for their support and loan of material. Thanks to Andrea Goharzadeh, Denis De Oliveira, Geoffrey Fournier, Pauline Salvatico and Julien Fezandelle for their help in the project. Thanks Pascal Mouquet for the satellite imageries. Finally, we thank \textit{Emlid} and \textit{Blue Robotic} for letting us use some of their images. 
This work is supported by the "IOT" project funded by FEDER INTERREG V and Prefet de La Réunion: grant \#20181306-0018039 and the Contrat de Convergence et de Transformation de la Préfecture de La Réunion.


\bibliographystyle{cas-model2-names}

\bibliography{global_biblio.bib}

\end{document}